\documentclass[12pt,a4paper]{article}
\usepackage{graphicx}
\usepackage{amsmath,amssymb}
\usepackage{amsfonts}
\usepackage{mathtools}
\usepackage{bm}

\usepackage{color}
\usepackage{enumitem}
\renewlist{description}{description}{10}
\setlistdepth{10}
\usepackage{lscape}
\usepackage{multirow}
\usepackage{mathrsfs}
\usepackage{cite}

\usepackage[height=8.85in,width=6.4in]{geometry}
\definecolor{BlueViolet}{rgb}{0.2, 0.00, 0.7}
\definecolor{Blue}{rgb}{0.15, 0.00, 0.9}
\usepackage[colorlinks=true,linkcolor=Blue,citecolor=Blue,urlcolor=BlueViolet]{hyperref}

\date{}

\begin{document}

\begin{titlepage}

\begin{center}

\vskip .55in

\begingroup
\centering

\hfill {\tt STUPP-23-264}
\vskip .55in

{\large\bf Model Building by Coset Space Dimensional Reduction Scheme Using Twelve-Dimensional Coset Spaces}

\endgroup

\vskip .4in

\renewcommand{\thefootnote}{\fnsymbol{footnote}}
{
Kento Asai$^{(a)}$\footnote{
  \href{mailto:kento@icrr.u-tokyo.ac.jp}
  {\tt kento@icrr.u-tokyo.ac.jp}},
Joe Sato$^{(b)}$\footnote{
  \href{mailto:sato-joe-mc@ynu.ac.jp}
  {\tt sato-joe-mc@ynu.ac.jp}}, 
Ryosuke Suda$^{(c)}$\footnote{
  \href{mailto:r.suda.813@ms.saitama-u.ac.jp}
  {\tt r.suda.813@ms.saitama-u.ac.jp}}, 
Yasutaka Takanishi$^{(c)}$\footnote{
  \href{mailto:sci56439@mail.saitama-u.ac.jp}
  {\tt sci56439@mail.saitama-u.ac.jp}}, \\
 and
Masaki J. S. Yang$^{(c)}$\footnote{
  \href{mailto:mjsyang@mail.saitama-u.ac.jp}
  {\tt mjsyang@mail.saitama-u.ac.jp}}
}

\vskip 0.2in

\begingroup\small
\begin{minipage}[t]{0.9\textwidth}
\centering\renewcommand{\arraystretch}{0.9}
{\it
\begin{tabular}{c@{\,}l}
$^{(a)}$
& Institute for Cosmic Ray Research (ICRR), The University of Tokyo, Kashiwa,\\
& Chiba 277--8582, Japan \\[2mm]
$^{(b)}$
& Department of Physics, Faculty of Engineering Science, Yokohama National University, \\
& Yokohama 240--8501, Japan \\[2mm]
$^{(c)}$
& Department of Physics, Faculty of Science, Saitama University, Saitama 338--8570, \\ 
& Japan \\ 
\end{tabular}
}
\end{minipage}
\endgroup

\end{center}

\vskip .4in
    \begin{abstract}
        We investigate the twelve-dimensional gauge-Higgs unification models with an eight-dimensional coset space. 
        For each model, we apply the coset space dimensional reduction procedure and examine the particle contents of the resulting four-dimensional theory.
        Then, some twelve-dimensional $\mathrm{SO}(18)$ gauge theories lead to models of the $\mathrm{SO}(10)\times \mathrm{U(1)}$ grand unified theory in four dimensions, where fermions of the Standard Model appear in multiple generations along with scalars that may break the electroweak symmetry.
        The representations of the obtained scalars and fermions are summarized.
    \end{abstract} 
    \thispagestyle{empty}
\end{titlepage}

\section{Introduction}\label{sec:intro}

The Standard Model (SM) of particle physics is a highly successful theory that serves as a language to precisely describe phenomena related to elementary particles, especially those observed in accelerator experiments. The discovery of the Higgs scalar in the year 
2012 fulfilled the last missing piece of the SM. This fact established 
the success of the SM and we are able to say that the SM is the standard model of elementary particle physics. However, unfortunately, the SM cannot predict the properties of the Higgs scalar, such as its mass and its coupling constant, and does not even refer to its theoretical origin. Therefore, a theory that explains the origin of the Higgs scalar beyond the SM has been intensely studied for many years.

One of the theories is called the gauge-Higgs unification (GHU) \cite{Manton:1979kb, Fairlie:1979at, Fairlie:1979zy} (for further developments, see Refs.~\cite{Hall:2001zb, Burdman:2002se, Scrucca:2003ra, Scrucca:2003ut, Haba:2004bh, Haba:2004qf, Hasegawa:2004zz,Hosotani:2006qp, Medina:2007hz, Sakamura:2007qz, Nomura:2008sx, Hosotani:2015hoa}) that explains the origin of the Higgs scalar.
In this framework, the theoretical origin of the Higgs scalar field is considered to be the extra-dimensional components of higher-dimensional ($D$-dimensional) gauge fields, and the properties of the Higgs scalar are attributed to the gauge symmetry and compactification scale of the higher-dimensional theory.
After the dimensional reduction, we can obtain a theory that contains the Higgs scalar with desired properties. One such possible dimensional reduction is the scheme of the coset space dimensional reduction (CSDR)\cite{Forgacs:1979zs, Kapetanakis:1992hf, Chatzistavrakidis:2007by, Douzas:2008va, Jittoh:2008jc, Jittoh:2008bs}, where the extra space is assumed to be a coset space $S/R$ of a compact Lie group $S$ by its subgroup $R\subset S$, and symmetry transformations of the extra space are identified with gauge transformations in higher dimensions. These constraints determine nicely the gauge group and particle content in the resulting four-dimensional theory.
Our main aim is to search for higher-dimensional models which
describe, after CSDR, chiral theories in four dimensions as in the SM, which means that there are 
left-handed and right-handed fermions that belong to different representations of the gauge group. 

The case of $D=4n+2$ dimensions ($n\in\mathbb{N}$) has been the most fascinating in this respect and has drawn much attention \cite{Chapline:1982wy,Wetterich:1982ed,Wetterich:1983ye}.
One can obtain a chiral theory in four dimensions after the dimensional reduction even if one chooses that vector-like fermions present in these dimensions.
However, no phenomenologically promising models have been found in $D=6,10, 14$ dimensions \cite{Chapline:1982wy, Wetterich:1983ye, Bais:1985yd, Pilch:1985pm, Kozimirov:1987ce, Farakos:1988ks, Kapetanakis:1990xx, Hanlon:1993aq}. 

It, therefore, becomes necessary to consider the case of $D=4n$, which is different from $D=4n+2$ because vector-like fermions in these dimensions cannot produce chiral fermions in four dimensions. Thus, one must introduce fermions belonging to complex representations of the gauge group in the higher dimensions. In our previous research \cite{Jittoh:2009th}, we have investigated the smallest dimension, i.e., $D=8$ and we have shown the possibility of building phenomenologically realistic models even for $D\neq 4n+2$. 
The purpose of the present study is to investigate the case of $D=12$, the next smallest dimension after $D=8$. Dimension twelve is also of interest from a purely theoretical perspective \cite{Vafa:1996xn, Mizoguchi:2014gva, Choi:2015gia} and is another motivation for our study.

In this paper, we investigate twelve-dimensional gauge theories with an eight-dimensional coset space as the extra dimension and explore the possibility of obtaining the SM, Grand Unified Theories (GUTs), and their analogues through coset space dimensional reduction. First, we enumerate possible coset spaces $S/R$ and gauge groups $G$ in twelve dimensions under plausible assumptions. 
Then the four-dimensional gauge groups $H$ arise from these settings. 
For each candidate, we introduce fermions and perform the dimensional reduction to obtain the resulting scalar and fermionic fields in four dimensions and determine their representations under the gauge group $H$.

This article is organized as follows. In Section \ref{sec:CSDR_review}, we review the coset space dimensional reduction. 
Next, in Section \ref{sec:search_for_candidates}, we classify models aiming for a twelve-dimensional theory that gives rise to phenomenologically desirable models in four dimensions. Finally, in Section \ref{sec:summary}, we summarize our results and discuss the implications of the models obtained in four dimensions.
\section{Coset space dimensional reduction in twelve dimensions}\label{sec:CSDR_review}
In this section, we review the scheme of the coset space dimensional reduction and discuss its specific application in $D=4+8$ dimensions \cite{Kapetanakis:1992hf}.

\subsection{Coset space dimensional reduction}
Let $S$ be a compact Lie group and $R \subset S$ be its compact subgroup. We denote their coset space by $S/R$. 
The group $S$ is the symmetry group of $S/R$ and $R$ is its isotropy group (also known as the little group).
The higher-dimensional coordinates are denoted by $X^M=(x^\mu,y^\alpha)$ for $M=0,\dots, D-1$, $\mu=0,1,2,3$, $\alpha = 4,\dots,D-1$. The tangent space of $S/R$ has a local symmetry group, $\mathrm{SO}(d)$, where $d$ is the dimension of the coset space, i.e. $d=\dim S/R$, which includes $R$ as a subgroup.
We consider a Yang-Mills theory on $\mathbb{R}^{1,3} \times S/R$, where $S/R$ is added as a direct product to the four-dimensional Minkowski space $\mathbb{R}^{1,3}$.
Furthermore, the gauge group of the theory is assumed to be a compact Lie group $G$.
We introduce a higher-dimensional fermion that belongs to a 
 representation $F$ of the gauge group $G$.
We then impose symmetry conditions on fields in the theory as in Refs.~\cite{Forgacs:1979zs, Witten:1976ck, Jackiw:1980jq, Olive:1982ai, Palla:1983re, Kubyshin:1987cv,Kubyshin:1989it}. 
These conditions require a transformation of the fields under 
a $S$ symmetry $s: y\mapsto y^s$ is canceled out by a gauge transformation $g$ of $G$.
As a result, the Lagrangian $\mathscr{L}^{(D)}(x,y)$ of the higher-dimensional theory loses its dependence on the extra-dimensional coordinates $y$.
Integrating out the extra dimensions gives the Lagrangian 
$\mathscr{L}^{(4)}(x)$ of the four-dimensional theory.
Moreover, the following representation-theoretic constraints are implied by CSDR to derive the content of the four-dimensional theory.

\begin{enumerate}[label=(\roman*)]
    \item The isotropy group $R$ of the coset space $S/R$ is a subgroup of $G$, which is denoted by $R_G \subset G$.
    \item The gauge group $H$ in four-dimensional theory is the centralizer of $R$ in $G$, i.e., the set of all elements in $G$ that commute with any element in $R$.
    \begin{align}
        H = C_G(R_G) = \{g\in G\mid gr=rg \quad (\forall r \in R)\}.
    \end{align}
    \item The representation of the gauge group $H$ for scalar fields in a four-dimensional theory is specified as follows. First, we decompose the adjoint representation $\operatorname{ad} S$ of the symmetry group of the coset space $S/R$ under the subgroup $R$:
    \begin{align}
        S &\supset R,\notag \\
        \operatorname{ad} S &= \operatorname{ad}R \oplus \left(\bigoplus_{i} \rho_i\right) ,\label{eq:ads_decomp}
    \end{align}
    where ${\rho_i}$ are irreducible representations of $R$.
    We also write the decomposition of the adjoint representation $\operatorname{ad} G$ of the gauge group $G$ under the action of $R_G \times H$ as follows. 
    \begin{align}
        G &\supset R_G\times H,\notag \\
        \operatorname{ad}G &= \bigoplus_{j}(r_j,h_j) \notag \\
        &= (\operatorname{ad}R,\bm{1})\oplus(\bm{1},\operatorname{ad}H)\oplus \cdots \, . \label{eq:adg_decomp}
    \end{align}
    Here, ${r_j}$ and ${h_j}$ are irreducible representations of $R$ and $H$, respectively.
    If equivalent irreducible $R$ representations $\rho_i$ and $r_j$ appear in the decompositions, Eq.~\eqref{eq:ads_decomp} and Eq.~\eqref{eq:adg_decomp}, the scalar fields belonging to $h_j$, appear in the four-dimensional theory.    
    \par In practice, it is useful to consider the $R$ decomposition of $\mathrm{SO}(d)$ vector, as described below.  
    The geometry of the coset space $S/R$ is governed by its symmetry group $S$ and its isotropy subgroup $R$.
    We write the generators of $S$ as $\{Q_A \} = \{Q_i,Q_a \}$, where $A=1,\dots,\dim S$, $i=1,\dots,\dim R$, and $a=1,\dots,d$ with $d = \dim S/R$. 
    $\{Q_i\}$ are the generators of $R$ and $\{Q_a\}$ are the rest of the generators (coset generators).
    It is important that the coset generators ${Q_a}$ span the tangent space of $S/R$, so they transform as vectors, denoted by $\bm{v}$, under the local symmetry $\mathrm{SO}(d)$ of the tangent space.
    Thus $S$, $R$, and $\mathrm{SO}(d)$ satisfy
        \begin{align}
            S\supset &R,\notag\\
            \operatorname{ad}S = &\operatorname{ad} R + \bm{v}. \label{eq:s_r_sod}
        \end{align}
    From this relation and Eq.~\eqref{eq:ads_decomp}, we see that the $H$ representations for the four-dimensional scalar fields can be obtained by comparing the two decompositions.
    One is $R$ decomposition of the $\mathrm{SO}(d)$ vector $\bm{v}$
    \begin{align}
        \mathrm{SO}(d) &\supset R,\notag \\
        \bm{v} &= \left(\bigoplus_i \rho_i\right),
    \end{align}
    and the other is the $R\times H$ decomposition of $\operatorname{ad}G$ shown in Eq.~\eqref{eq:adg_decomp}.
    \item The representations of the gauge group $H$ for spinor fields in four-dimensional theory are derived according to the following procedure, as in Ref.~\cite{Manton:1981es}.
    First, we decompose the spinor representation $\sigma_d$ of the local symmetry group $\mathrm{SO}(d)$, under the subgroup $R$,
    \begin{align}
        \mathrm{SO}(d) &\supset R,\notag \\
        \sigma_d &= \bigoplus_i \sigma_i. \label{eq:spinor_decomp}
    \end{align}
    Here, ${\sigma_i}$ are irreducible representations of $R$.
    The higher-dimensional spinor field belongs to the representation $F$ of the gauge group $G$, as we have already mentioned.
    We write its $R_G\times H$ decomposition as
    \begin{align}
        G &\supset R_G\times H,\notag \\
        F &= \bigoplus_{j}(r_j,h_j), \label{eq:F_decomp}
    \end{align}
    where ${r_j}$ and ${h_j}$ are irreducible representations of $R$ and $H$, respectively.
    If there are equivalent irreducible $R$ representations $\sigma_i$ and $r_j$, the spinor fields in the irreducible representation $h_j$ appear in the four-dimensional theory.

\end{enumerate}

To obtain a theory with chiral spinor fields, the following conditions should be imposed in addition to the rules above.

\begin{enumerate}
    \item  The rank of $S,R$ must be the same: $\operatorname{rank}R=\operatorname{rank}S$. 
    This is a geometric requirement based on a non-trivial result for the Dirac operator on the coset space $S/R$ \cite{Bott:1965dc}.
    \item
    When $D=4n$, the spinors introduced in higher-dimensional theories must be Weyl spinors, and in this case, the representation $F$ of the gauge group $G$ to which they belong must be a complex representation.
    Furthermore, the isotropy group $R$ of the coset space $S/R$ must also admit a complex representation \cite{Kapetanakis:1992hf}.
\end{enumerate}

\subsection{Application and remarks of CSDR on eight-dimensional coset spaces}

For the case of $d=8$ dimensional coset space, representations of the local $\mathrm{SO}(8)$ symmetry become important.
The $\mathrm{SO}(8)$ group has three distinct self-conjugate representations, which are denoted as $\bm{8}_\text{v}$, $\bm{8}_\text{c}$, and $\bm{8}_\text{s}$. 
The representation $\bm{8}_\text{v}$ is the usual real eight-dimensional vector representation, while $\bm{8}_\text{c}$ and $\bm{8}_\text{s}$ are both real eight-dimensional spinor representations that describe Weyl spinors. 
We adopt the convention that $\bm{8}_\text{c}$ has positive chirality and $\bm{8}_\text{s}$ has negative chirality. 
The three eight-dimensional representations are related to each other by the outer automorphisms of $\mathrm{SO}(8)$ and in some contexts the vector $\bm{8}_\text{v}$ and the spinors $\bm{8}_\text{c}$ and $\bm{8}_\text{s}$ can be considered equivalent. 
For example, it is possible that the decompositions of $\bm{8}_\text{v}, \bm{8}_\text{c}$, and $\bm{8}_\text{s}$ under a subgroup of $\mathrm{SO}(8)$ may be interchanged for different choices of embedding.
This equivalence of the three eight-dimensional representations is called triality (see ``fun with $\mathrm{SO}(8)$'' in Ref.~\cite{Georgi:1982jb}). 
We will now discuss some important notes regarding CSDR over the eight-dimensional coset space, using the representations $\bm{8}_\text{v}$, $\bm{8}_\text{c}$, and $\bm{8}_\text{s}$.

First, let us discuss scalars.
A scalar field in four dimensions arises from the extra-dimensional components $(\phi_a)$ of a twelve-dimensional $\mathrm{SO}(1,11)$ vector field $A = (A_M) = (A_\mu,\phi_a)$, where $a=1,\dots,8$. 
A twelve-dimensional vector is decomposed under the subgroup $\mathrm{SO}(1,3)\times \mathrm{SO}(8)$ as
\begin{align}
\mathrm{SO}(1,11)&\supset \mathrm{SO}(1,3)\times \mathrm{SO}(8), \notag \\
\bm{12} &= (\bm{4},\bm{1}) + (\bm{1},\bm{8}_\text{v}).
\end{align}
Since $\phi=(\phi_a)$ behaves as a vector $\bm{8}_\text{v}$ under $\mathrm{SO}(8)$, the representation of the gauge group $H$ to which the four-dimensional scalar field belongs can be obtained by comparing the decomposition of $\bm{8}_\text{v}$ under $R$,
\begin{align}
    \mathrm{SO}(8) &\supset R,\notag \\
    \bm{8}_\text{v} &= \left(\bigoplus_i \rho_i\right),\label{eq:matching_scalar}
\end{align}
and the decomposition of $\operatorname{ad}G$ under $R\times H$ (see Eq.~\eqref{eq:adg_decomp}).

Next, we discuss spinors.
As mentioned in the previous section, the Weyl spinors must be introduced in $D=12$ dimensions.
We denote the $\mathrm{SO}(1,11)$ Weyl spinor with positive chirality as $\bm{32}$ and negative chirality as $\bm{32}^\prime$, respectively.
They transform under the subgroup $\mathrm{SO}(1,3)\times \mathrm{SO}(8)$ as follows:
\begin{align}
    \mathrm{SO}(1,11) &\supset \mathrm{SO}(1,3)\times \mathrm{SO}(8),\notag \\
    \bm{32} &= (\bm{2},\bm{1};\bm{8}_\text{s}) + (\bm{1,2};\bm{8}_\text{c}), \notag \\
    \bm{32}^\prime &= (\bm{2},\bm{1};\bm{8}_\text{c}) + (\bm{1},\bm{2};\bm{8}_\text{s}),
\end{align}
where $(\bm{1},\bm{2})$ and $(\bm{2},\bm{1})$ represent right-handed and left-handed Weyl spinors with positive and negative chirality in four dimensions, respectively.
If $\bm{32}$ Weyl spinor is introduced in twelve dimensions, the $H$ representation of the surviving right-handed spinor in four dimensions is derived by matching the representations that appear in the $R$ decomposition in Eq.~\eqref{eq:spinor_decomp} of $\sigma_d = \bm{8}_\text{c}$ and in the $R\times H$ decomposition of the $G$ representation $F$ in Eq.~\eqref{eq:F_decomp}.
Introducing $\bm{32}^\prime$ in twelve dimensions yields right-handed spinors from $\bm{8}_\text{s}$ and left-handed spinors from $\bm{8}_\text{c}$.
In this paper, we introduce $\bm{32}$ as the Weyl spinor.

\begin{table}[thp]
    \centering
    \caption{The list of eight-dimensional coset spaces $S/R$ that satisfy $\operatorname{rank}R=\operatorname{rank}S$. The names of the coset spaces are followed by ``max'' or ``non-max'' to indicate whether $R$ is a maximal subgroup of $S$ or not. The column ``$\mathrm{U}(1)$'s in $R$'' indicates the number of $\mathrm{U}(1)$ factors in $R$, and the column ``$R$ complex reps'' indicates whether $R$ has complex representations (``Yes'') or not (``No'').}
    \label{tab:list_8dim_coset_modified}
    \scalebox{0.8}{
    \begin{tabular}{cccc}
       \hline
        Coset space                                                                    & dimension      & $\mathrm{U}(1)$'s in $R$ & $R$ complex reps \\ \hline
        $\mathrm{SU}(5)/\mathrm{SU}(4)\times \mathrm{U}(1)             $                                  & 8       & 1               & Yes        \\
        $\mathrm{SO}(9)/\mathrm{SO}(8)                 $                                         & 8       & 0               & No        \\
        $\mathrm{Sp}(6)/\mathrm{SU}(2)\times \mathrm{Sp}(4)            $                                  & 8       & 0               & No        \\
        $\mathrm{G}_2/\mathrm{SU}(2)\times \mathrm{SU}(2)               $                                 & 8       & 0               & No        \\
        ${\mathrm{Sp}(4)/\mathrm{U}(1)\times \mathrm{U}(1)}_{\text{non-max}}       $                      & 8       & 2               & Yes        \\
        ${\mathrm{SU}(4)/\mathrm{SU}(2)\times \mathrm{SU}(2)\times \mathrm{U}(1)}_{\text{non-max}} $               & 8       & 1               & Yes        \\ \hline
        $(\mathrm{SU}(4)/\mathrm{SU}(3)\times \mathrm{U}(1))\times (\mathrm{SU}(2)/\mathrm{U}(1))             $             & 6+2     & 2               & Yes        \\
        $(\mathrm{SO}(7)/\mathrm{SO}(6))\times (\mathrm{SU}(2)/\mathrm{U}(1))                 $                    & 6+2     & 1               & Yes        \\
        $(\mathrm{G}_2/\mathrm{SU}(3))\times (\mathrm{SU}(2)/\mathrm{U}(1))                    $                   & 6+2     & 1               & Yes        \\
        $({\mathrm{SU}(3)/\mathrm{U}(1)\times \mathrm{U}(1)}_{\text{non-max}} )\times (\mathrm{SU}(2)/\mathrm{U}(1))      $ & 6+2     & 3               & Yes        \\
        $({\mathrm{Sp}(4)/\mathrm{SU}(2)\times \mathrm{U}(1)}_{\text{max}} )\times (\mathrm{SU}(2)/\mathrm{U}(1))         $ & 6+2     & 2               & Yes        \\
        $({\mathrm{Sp}(4)/\mathrm{SU}(2)\times \mathrm{U}(1)}_{\text{non-max}})\times (\mathrm{SU}(2)/\mathrm{U}(1))     $  & 6+2     & 2               & Yes        \\ \hline
        $(\mathrm{SU}(3)/\mathrm{SU}(2)\times \mathrm{U}(1))^2             $                              & 4+4     & 2               & Yes        \\
        $(\mathrm{Sp}(4)/\mathrm{SU}(2)\times \mathrm{SU}(2))\times (\mathrm{SU}(3)/\mathrm{SU}(2)\times \mathrm{U}(1))          $   & 4+4     & 1               & Yes        \\
        $(\mathrm{Sp}(4)/\mathrm{SU}(2)\times \mathrm{SU}(2))^2            $                              & 4+4     & 0               & No        \\ \hline
        $(\mathrm{SU}(3)/\mathrm{SU}(2)\times \mathrm{U}(1))\times(\mathrm{SU}(2)/\mathrm{U}(1))^2             $            & 4+2+2   & 3               & Yes        \\
        $(\mathrm{Sp}(4)/\mathrm{SU}(2)\times \mathrm{SU}(2))\times(\mathrm{SU}(2)/\mathrm{U}(1))^2            $            & 4+2+2   & 2               & Yes        \\ \hline
        $(\mathrm{SU}(2)/\mathrm{U}(1))^4                  $                                     & 2+2+2+2 & 4               & Yes        \\ \hline
    \end{tabular}
    }
\end{table}


\section{Search for candidate models}\label{sec:search_for_candidates}

In this section, we explore realistic models in CSDR using the eight-dimensional coset spaces.
First, we list in Table \ref{tab:list_8dim_coset_modified} the eight-dimensional coset spaces satisfying the condition~: $\operatorname{rank}R=\operatorname{rank}S$.
As an ansatz for model building, the gauge group $H$ in four dimensions should be the SM gauge group $G_\text{SM}=\mathrm{SU}(3)\times \mathrm{SU}(2)\times \mathrm{U}(1)$, as well as the gauge groups $\mathrm{SU}(5)$, $\mathrm{SO}(10)$, and $\mathrm{E}_6$ used in grand unified theories, and those with up to one $\mathrm{U}(1)$ factor.
Thus, among the coset spaces listed in Table \ref{tab:list_8dim_coset_modified}, those with two or more $\mathrm{U}(1)$ factors in $R$ are excluded from our search.
The reason for this assumption is that the $\mathrm{U}(1)$ factors of $R$ themselves enter the centralizer of $R$.
This implies that the more $\mathrm{U}(1)$ factors are included in $R$, the more $\mathrm{U}(1)$ factors there are in $H$, leading to additional complexity from taking linear combinations of $\mathrm{U}(1)$ generators.
Considering these conditions, the coset spaces $S/R$ to be examined in this study are summarized in Table \ref{tab:list_8dim_coset_rearranged}.

\begin{table}[htbp]
\centering
\caption{The list of the eight-dimensional coset spaces $S/R$ that we are investigating. The columns labeled ``vector/spinors under $R$'' indicate how the eight-dimensional representation of $\mathrm{SO}(8)$ decomposes under $R$ for each $S/R$.}
\label{tab:list_8dim_coset_rearranged}
\scalebox{0.68}{
\begin{tabular}{c|l|l}
\hline
$S/R$                                                                                      & vector under $R$                 & spinors under $R$                                                                                  \\ \hline
$\mathrm{SU}(5)/\mathrm{SU}(4)\times \mathrm{U}(1)$                                                                   & $\bm{8}_\text{v} = \bm{4}(1)+\overline{\bm{4}}(-1)$     & \begin{tabular}[l]{@{}l@{}}\raisebox{-1mm}{$\bm{8}_\text{c}=\bm{4}(-1)+\overline{\bm{4}}(1)$}\\ $\bm{8}_\text{s}=\bm{6}(0)+\bm{1}(2)+\bm{1}(-2)$\end{tabular} \\ \hline
$\mathrm{SU}(4)/\mathrm{SU}(2)\times \mathrm{SU}(2)\times \mathrm{U}(1)$                                                       & $\bm{8}_\text{v}=(\bm{2},\bm{2})(1)+(\bm{2},\bm{2})(-1)$ & \begin{tabular}[l]{@{}l@{}}$\bm{8}_\text{c}=(\bm{3},\bm{1})(0)+(\bm{1},\bm{3})(0)+(\bm{1},\bm{1})(2)+(\bm{1},\bm{1})(-2)$\\ $\bm{8}_\text{s}=(\bm{2},\bm{2})(1)+(\bm{2},\bm{2})(-1)$\end{tabular}                  \\ \hline
\begin{tabular}[c]{@{}c@{}}$\mathrm{SO}(7)/\mathrm{SO}(6)$ \\ $\times$\\$ 
 \mathrm{SU}(2)/\mathrm{U}(1)$ \end{tabular}     & $\bm{8}_\text{v}=\bm{6}(0)+\bm{1}(2)+\bm{1}(-2)$  & \begin{tabular}[c]{@{}c@{}}$\bm{8}_\text{c}=\bm{4}(1)+\overline{\bm{4}}(-1)$\\ $\bm{8}_\text{s}=\bm{4}(-1)+\overline{\bm{4}}(1)$\end{tabular}                                                                                          \\ \hline
\begin{tabular}[c]{@{}c@{}}$ \mathrm{G}_2/\mathrm{SU}(3)$\\ $\times$\\$ \mathrm{SU}(2)/\mathrm{U}(1)$\end{tabular}           & $\bm{8}_\text{v} = \bm{3}(0)+\overline{\bm{3}}(0)+\bm{1}(2)+\bm{1}(-2)$ & \begin{tabular}[c]{@{}c@{}}$\bm{8}_\text{c}=\bm{3}(-1)+\overline{\bm{3}}(1) + \bm{1}(-1) + \bm{1}(1)$\\ $\bm{8}_\text{s}=\bm{3}(1)+\overline{\bm{3}}(-1)+\bm{1}(1)+\bm{1}(-1)$\end{tabular} \\ \hline
\begin{tabular}[c]{@{}c@{}}$\mathrm{Sp}(4)/\mathrm{SU}(2)\times \mathrm{SU}(2)$\\ $\times $\\$\mathrm{SU}(3)/\mathrm{SU}(2)\times \mathrm{U}(1)$\end{tabular} & $\bm{8}_\text{v}=(\bm{2},\bm{1},\bm{1})(2)+(\bm{1},\bm{2},\bm{2})(0)+(\bm{2},\bm{1},\bm{1})(-2)$  & \begin{tabular}[c]{@{}c@{}}$\bm{8}_\text{c}=(\bm{1},\bm{2},\bm{1})(2) + (\bm{2},\bm{1},\bm{2})(0) + (\bm{1},\bm{2},\bm{1})(-2)$\\ $\bm{8}_\text{s}=(\bm{1},\bm{1},\bm{2})(2) + (\bm{2},\bm{2},\bm{1})(0) + (\bm{1},\bm{1},\bm{2})(-2)$\end{tabular} \\ \hline
\end{tabular}}
\end{table}

We examine the twelve-dimensional gauge groups $G$ such that the desired gauge group $H$ is obtained for the coset space $S/R$. 
This is achieved by considering a sequence of maximal regular subgroups of $G$.
By identifying $R$ in the sequence, we determine the embedding $R_G\subset G$, and then the other factors in $G$ times $\mathrm{U}(1)$ factors give $H$.
We do not consider non-regular maximal subgroups (called special subgroups) to embed $R$ into $G$, because we need a much larger $G$ to embed $R$ as a special subgroup than as a regular subgroup.
Thus we consider sequences of maximal regular subgroups of $G$ \cite{Yamatsu:2017mei}.

In Table \ref{tab:list_12dim_gauge_group}, we present the candidates for the twelve-dimensional gauge group $G$ and the resulting four-dimensional gauge groups $H$.
Note that there is no $G$ that gives rise to $H=\mathrm{E}_6$ in four dimensions because $G$ in $D=12$ is restricted to groups with complex representations, which cannot contain $\mathrm{E}_6$ as a subgroup.

\begin{table}[tbp]
\centering
\caption{The list of candidate twelve-dimensional gauge groups $G$ and the resulting four-dimensional gauge groups $H$ for the coset space $S/R$ under investigation. }
\label{tab:list_12dim_gauge_group}
\scalebox{0.8}{
\begin{tabular}{c|c||c}
\hline
$S/R$                                                                               & $G        $ & $H                                       $ \\ \hline
\multirow{3}{*}{$\mathrm{SU}(5)/\mathrm{SU}(4)\times \mathrm{U}(1)$}                                           & $\mathrm{SU}(9)    $ & $\mathrm{SU}(3)\times \mathrm{SU}(2)\times \mathrm{U}(1)\times \mathrm{U}(1) $ \\
                                                                                    & $\mathrm{SU}(9)    $ & $\mathrm{SU}(5)\times \mathrm{U}(1)                        $ \\
                                                                                    & $\mathrm{SO}(18)   $ & $\mathrm{SO}(10)\times \mathrm{U}(1)                       $ \\ \hline
$\mathrm{SU}(4)/\mathrm{SU}(2)\times \mathrm{SU}(2)\times \mathrm{U}(1)$                                                & $\mathrm{SO}(14)   $ & $\mathrm{SU}(5)\times \mathrm{U}(1)                        $ \\ \hline
\multirow{3}{*}{${[}\mathrm{SO}(7)/\mathrm{SO}(6){]}\times{[}\mathrm{SU}(2)/\mathrm{U}(1){]}$}                          & $\mathrm{SU}(9)    $ & $\mathrm{SU}(3)\times \mathrm{SU}(2)\times \mathrm{U}(1)\times \mathrm{U}(1) $ \\
                                                                                    & $\mathrm{SU}(9)    $ & $\mathrm{SU}(5)\times \mathrm{U}(1)                        $ \\
                                                                                    & $\mathrm{SO}(18)   $ & $\mathrm{SO}(10)\times \mathrm{U}(1)                       $ \\ \hline
\multirow{3}{*}{${[}\mathrm{G}_2/\mathrm{SU}(3){]}\times {[}\mathrm{SU}(2)/\mathrm{U}(1){]}$}                           & $\mathrm{E}_6      $ & $\mathrm{SU}(3)\times \mathrm{SU}(2)\times \mathrm{U}(1)            $ \\
                                                                                    & $\mathrm{SU}(8)    $ & $\mathrm{SU}(3)\times \mathrm{SU}(2)\times \mathrm{U}(1)\times \mathrm{U}(1) $ \\
                                                                                    & $\mathrm{SU}(8)    $ & $\mathrm{SU}(5)\times \mathrm{U}(1)                        $ \\ \hline
\multirow{2}{*}{${[}\mathrm{Sp}(4)/\mathrm{SU}(2)\times \mathrm{SU}(2){]}\times {[}\mathrm{SU}(3)/\mathrm{SU}(2)\times \mathrm{U}(1){]}$} & $\mathrm{SO}(14)   $ & $\mathrm{SU}(3)\times \mathrm{SU}(2)\times \mathrm{U}(1)            $ \\
                                                                                    & $\mathrm{SO}(18)   $ & $\mathrm{SO}(10)\times \mathrm{U}(1)                       $ \\ \hline 
\end{tabular}
}
\end{table}

Next, we consider the $G$ representations $F$ of higher-dimensional spinors. 
The dimension of $F$ is set to be less than $1000$ to avoid the appearance of many unnecessary fermions in four dimensions.
We then perform CSDR for all possible combinations of $(S/R, G, F)$ and search for cases where both scalars and at least one generation of fermions in the SM are obtained in four dimensions.
The analysis gives three interesting twelve-dimensional models, as shown in Table \ref{tab:obtained_models}.

\vspace*{1mm}

\begin{table}[htbp]
\centering
\caption{The inputs to CSDR, including the coset space $S/R$, the twelve-dimensional gauge group $G$, and the $G$ representation $F$ of twelve-dimensional fermions, along with the resulting content of the four-dimensional model (the gauge group $H$, the scalar representations, and the fermion representations).}
\label{tab:obtained_models}
\scalebox{0.78}{\begin{tabular}{c|c|c||c|c|c}
\hline
$S/R$ & $G$ &  $F$ & $H$ & scalars & fermions \\ \hline
$\mathrm{SU}(5)/\mathrm{SU}(4)\times \mathrm{U}(1)$ & $\mathrm{SO}(18)$ & $\bm{256}$ & $\mathrm{SO}(10)\times \mathrm{U}(1)$  & $\bm{10}(1)+\overline{\bm{10}}(-1)$  & \begin{tabular}[c]{@{}c@{}}$\bm{16}(2)+\bm{16}(0)+\bm{16}(-2)$\\ $+\bm{16}(1)+\bm{16}(-1)$\end{tabular}       \\ \hline
\begin{tabular}[c]{@{}c@{}}$\mathrm{SO}(7)/\mathrm{SO}(6)$\\ $\times $\\ $\mathrm{SU}(2)/\mathrm{U}(1)$\end{tabular}           & $\mathrm{SO}(18)$ & $\bm{256}$ & $\mathrm{SO}(10)\times \mathrm{U}(1)$ & $\bm{10}(2)+\bm{10}(0)+\bm{10}(-2)$ & \begin{tabular}[c]{@{}c@{}}$\bm{16}(1)+\bm{16}(-1)$\\ $+\bm{16}(1)+\bm{16}(-1)$\end{tabular}             \\\hline
\begin{tabular}[c]{@{}c@{}}$\mathrm{Sp}(4)/\mathrm{SU}(2)\times \mathrm{SU}(2)$\\ $\times $\\ $\mathrm{SU}(3)/\mathrm{SU}(2)\times \mathrm{U}(1)$\end{tabular} & $\mathrm{SO}(18)$ & $\bm{256}$ & $\mathrm{SO}(10)\times \mathrm{U}(1)$ & $\bm{10}(2)+\bm{10}(0)+\bm{10}(-2)$ & \begin{tabular}[c]{@{}c@{}}$\bm{16}(2)+\bm{16}(0)+\bm{16}(-2)$\\  $+\bm{16}(2)+\bm{16}(0)+\bm{16}(-2)$\end{tabular} \\ \hline
\end{tabular}}
\end{table}
For the models with $H=G_\text{SM}(\times \mathrm{U}(1)), \mathrm{SU}(5)\times \mathrm{U}(1)$, no candidates for twelve-dimensional models are found quite promising because the particle content of the obtained four-dimensional theory is not enough to reproduce the SM.
We describe the three models obtained in the following subsections.

\subsection{\texorpdfstring{$S/R = \mathrm{SU}(5)/\mathrm{SU}(4)\times \mathrm{U}(1), G=\mathrm{SO}(18),F=\bm{256}$}{S/R = SU(5)/R=SU(4)*U(1),G=\mathrm{SO}(18),F=256,H=\mathrm{SO}(10)*U(1)}}

Let us first consider the $H=\mathrm{SO}(10)\times \mathrm{U}(1)$ model obtained for the case $S/R=\mathrm{SU}(5)/\mathrm{SU}(4)\times \mathrm{U}(1)$, with $G=\mathrm{SO}(18)$ and $F=\bm{256}$.
For the embedding of $R=\mathrm{SU}(4)\times \mathrm{U}(1)$ into $G=\mathrm{SO}(18)$, we use the following subgroup decomposition:
\begin{align}
    \mathrm{SO}(18)
        & \supset \mathrm{SO}(10)\times \mathrm{SO}(8)_G\notag \\
        & \supset \mathrm{SO}(10)\times \mathrm{SU}(4)\times \mathrm{U}(1), \notag \\
    \operatorname{ad}\mathrm{SO}(18) = \bm{153} 
        & = (\bm{45},\bm{1}) + (\bm{1},\bm{28}) + (\bm{10},\bm{8}_\text{v}) \notag \\
        &= \underbrace{(\bm{45},\bm{1})(0)}_{\operatorname{ad}\mathrm{SO}(10)}  + \underbrace{(\bm{1},\bm{15})(0) + (\bm{1},\bm{1})(0)}_{\operatorname{ad}\mathrm{SU}(4)\times \mathrm{U}(1)}+ (\bm{1},\bm{6})(2) + (\bm{1},\bm{6})(-2) + (\bm{10},\bm{8}_\text{v}) .\label{eq:SO18supSU4U1}
\end{align}
Here, $\mathrm{SO}(8)_G\subset G$ represents the $\mathrm{SO}(8)$ subgroup in $G$, and $\mathrm{SO}(8)_\text{loc}$ is the $\mathrm{SO}(8)$ local symmetry of the tangent space of $S/R$. 
The embedding of $R$ into $G$ is specified by the choice for embedding $R$ into $\mathrm{SO}(8)_G$, which has an ambiguity due to the fact that the three eight-dimensional representations of $\mathrm{SO}(8)$ behave differently under $R$. 
On the other hand, the embedding of $R$ into $\mathrm{SO}(8)_\text{loc}$ is fixed as
\begin{align}
    \mathrm{SO}(8)_\text{loc}&\supset R=\mathrm{SU}(4)\times \mathrm{U}(1), \notag \\
    \bm{8}_\text{v} &= \bm{4}(1) + \overline{\bm{4}}(-1).\label{eq:8v_of_so8_to_su4u1}
\end{align}
To obtain a four-dimensional scalar using the CSDR rule in Eq.~\eqref{eq:matching_scalar}, we need to embed $R$ into $\mathrm{SO}(8)_G$ in the same way as we embed $R$ into $\mathrm{SO}(8)_\text{loc}$ in Eq.~\eqref{eq:8v_of_so8_to_su4u1}.
Such an embedding, given by
\begin{align}
    \mathrm{SO}(8)_G & \supset R=\mathrm{SU}(4)\times \mathrm{U}(1), \notag \\
    \bm{8}_\text{v} &= \bm{4}(1) + \overline{\bm{4}}(-1),
\end{align}
leads to the decomposition for the adjoint representation $\bm{153}$ of $G=\mathrm{SO}(18)$ under $R\times H=\mathrm{SO}(10)\times \mathrm{SU}(4)\times \mathrm{U}(1)$ in Eq.~\eqref{eq:SO18supSU4U1} as
\begin{align}
    \mathrm{SO}(18) \supset & ~ \mathrm{SO}(10)\times \mathrm{SU}(4)\times \mathrm{U}(1), \notag \\
    \bm{153} = &~(\bm{45},\bm{1})(0) + (\bm{1},\bm{15})(0) + (\bm{1},\bm{1})(0) + (\bm{1},\bm{6})(2) + (\bm{1},\bm{6})(-2) \notag \\ &+ (\bm{10},\bm{4})(1) + (\bm{10},\overline{\bm4})(-1).
\end{align}
Thus, comparing this and the decomposition of $\mathrm{SO}(8)_\text{loc}$ vector $\bm{8}_\text{v}$ under $R$ in Eq.~\eqref{eq:8v_of_so8_to_su4u1}, we obtain $\bm{10}(1)$ and $\bm{10}(-1)$ of $H=\mathrm{SO}(10)\times \mathrm{U}(1)$ as the four-dimensional scalar fields.

Furthermore, the $R$ decompositions of the $\mathrm{SO}(8)_\text{loc}$ spinors $\bm{8}_\text{c}$ and $\bm{8}_\text{s}$ in this case are given by:
\begin{align}
    \mathrm{SO}(8)_\text{loc} \supset & ~ \mathrm{SU}(4)\times \mathrm{U}(1),\notag \\
    \bm{8}_\text{c} = & ~ \bm{4}(-1) + \overline{\bm{4}}(1),\notag \\
    \bm{8}_\text{s} = & ~ \bm{6}(0) + \bm{1}(2) + \bm{1}(-2).
\end{align}
Comparing this and  the $R\times H$ decomposition of $F=\bm{256}$ of $G=\mathrm{SO}(18)$,
\begin{align}
    \mathrm{SO}(18) \supset & ~ \mathrm{SO}(10)\times \mathrm{SU}(4)\times \mathrm{U}(1),\notag \\
    \bm{256} = &~(\bm{16}, 1)(-2) + (\bm{16}, \bm{1})(2) + (\bm{16}, \bm{6})(0) \notag \\
    &+ (\overline{\bm{16}},\overline{\bm{4}})(1) + (\overline{\bm{16}},\bm{4})(-1),
\end{align}
we obtain four-dimensional left-handed spinor fields in $\bm{16}(2),\bm{16}(0),\bm{16}(-2)$ of $\mathrm{SO}(10)\times \mathrm{U}(1)$, while the right-handed spinors in $\overline{\bm{16}}(1),\overline{\bm{16}}(-1)$ also appear in four dimensions. 
Taking the charge conjugation of the right-handed spinor fields yields five $\bm{16}$'s, each of which contains the fermions of one generation of the SM.

\subsection{\texorpdfstring{$S/R = [\mathrm{SO}(7)/\mathrm{SO}(6)]\times [\mathrm{SU}(2)/\mathrm{U}(1)], G=\mathrm{SO}(18),F=\bm{256}$}{S/R = [SO(7)/SO(6)]*[SU(2)/U(1)],G=SO(18),F=256,H=SO(10)*U(1)}}

Next, we discuss the $H=\mathrm{SO}(10)\times \mathrm{U}(1)$ model obtained when the extra space is $S/R=[\mathrm{SO}(7)/\mathrm{SO}(6)]\times [\mathrm{SU}(2)/\mathrm{U}(1)]$ and $G=\mathrm{SO}(18)$ with $F=\bm{256}$.
Due to the Lie algebra isomorphism $\mathrm{SO}(6)\sim \mathrm{SU}(4)$, the embedding of $R=\mathrm{SO}(6)\times \mathrm{U}(1)$ into $G=\mathrm{SO}(18)$ can be identified by choosing $R=\mathrm{SO}(6)\times \mathrm{U}(1)\subset \mathrm{SO}(8)_G$ in the same way as $R\subset \mathrm{SO}(8)_\text{loc}$, which we have done in the previous section.
Since the $\mathrm{SO}(8)_\text{loc}$ vector transforms under $R$ as
\begin{align}
    \mathrm{SO}(8)_\text{loc} \supset & ~ \mathrm{SO}(6)\times \mathrm{U}(1), \notag \\
    \bm{8}_\text{v} = & ~ \bm{6}(0) + \bm{1}(2) + \bm{1}(-2), \label{eq:8v_of_so8_to_so6u1}
\end{align}
the embedding $R\subset \mathrm{SO}(8)_G$ is specified as
\begin{align}
    \mathrm{SO}(18) \supset & ~ \mathrm{SO}(10)\times \mathrm{SO}(6)\times \mathrm{U}(1), \notag \\
    \operatorname{ad}\mathrm{SO}(18) = \bm{153} = & ~ (\bm{45},\bm{1})(0) + (\bm{1},\bm{15})(0) + (\bm{1},\bm{1})(0) +(\bm{1},\bm{6})(2) + (\bm{1},\bm{6})(-2) \notag \\
    &+(\bm{10},\bm{6})(0) + (\bm{10},\bm{1})(2) + (\bm{10},\bm{1})(-2).
\end{align}
Comparing these two decompositions gives $\bm{10}(0),\bm{10}(2),\bm{10}(-2)$ of $\mathrm{SO}(10)\times \mathrm{U}(1)$ as the four-dimensional scalars.

As for the spinors, the decompositons of the $\mathrm{SO}(8)_\text{loc}$ spinors under $R$ are given by
\begin{align}
    \mathrm{SO}(8)_\text{loc} \supset & ~ \mathrm{SO}(6)\times \mathrm{U}(1),\notag \\
    \bm{8}_\text{c} = & ~ \bm{4}(1) + \overline{\bm{4}}(-1) ,\notag \\
    \bm{8}_\text{s} = & ~\bm{4}(-1) + \overline{\bm{4}}(1).
\end{align}
If this is compared to the decomposition of $F=\bm{256}$ of $G=\mathrm{SO}(18)$ under $R\times H$,
\begin{align}
    \mathrm{SO}(18) \supset & ~ \mathrm{SO}(10)\times \mathrm{SO}(6)\times \mathrm{U}(1),\notag \\
    \bm{256} = & ~ (\bm{16},\bm{4})(-1) + (\bm{16},\overline{\bm{4}})(1) 
    + (\overline{\bm{16}},\bm{4})(1) + (\overline{\bm{16}},\overline{\bm{4}})(-1),
\end{align}
we obtain the four-dimensional left-handed spinors in $\bm{16}(\pm 1)$ of $\mathrm{SO}(10)\times \mathrm{U}(1)$ and the right-handed spinors in $\overline{\bm{16}}(\pm 1)$.
In other words, four generations of fermions in $\bm{16}$ of $\mathrm{SO}(10)$ appear in four-dimensional theory if we take the charge conjugation of the right-handed spinors.

\subsection{\texorpdfstring{$[\mathrm{Sp}(4)/\mathrm{SU}(2)\times \mathrm{SU}(2)]\times [\mathrm{SU}(3)/\mathrm{SU}(2)\times \mathrm{U}(1)]$}{S/R = [Sp(4)/SU(2)*SU(2)]*[SU(3)/SU(2)*U(1)],G=SO(18),F=256,H=SO(10)*U(1)}}

Finally, we consider the case of the $H=\mathrm{SO}(10)\times \mathrm{U}(1)$ model obtained from $G=\mathrm{SO}(18)$ and $F=\bm{256}$ with the coset space $S/R=[\mathrm{Sp}(4)/\mathrm{SU}(2)\times \mathrm{SU}(2)]\times [\mathrm{SU}(3)/\mathrm{SU}(2)\times \mathrm{U}(1)]$.
The embedding of $R=\mathrm{SU}(2)\times \mathrm{SU}(2)\times \mathrm{SU}(2)\times \mathrm{U}(1)$ into $G=\mathrm{SO}(18)$ is also provided by adjusting the embedding $R\subset \mathrm{SO}(8)_G$ to match $R \subset \mathrm{SO}(8)_\text{loc}$:
\begin{align}
    \mathrm{SO}(8)_\text{loc} \supset & ~ \mathrm{SU}(2)\times \mathrm{SU}(2)\times \mathrm{SU}(2)\times \mathrm{U}(1),\notag \\
    \bm{8}_\text{v} =&~ (\bm{2},\bm{1},\bm{1})(2)+(\bm{1},\bm{2},\bm{2})(0)+(\bm{2},\bm{1},\bm{1})(-2). \label{eq:8v_of_so8_to_2221}
\end{align}
Then, $R\subset \mathrm{SO}(8)_G$ is chosen to be the same as above, which gives the $R\times H$ decomposition of the adjoint representation $\bm{153}$ of $G=\mathrm{SO}(18)$ as
\begin{align}
    \mathrm{SO}(18) \supset & ~ \mathrm{SO}(10) \times \mathrm{SU}(2)\times \mathrm{SU}(2)\times \mathrm{SU}(2)\times \mathrm{U}(1),\notag \\
     153 = & ~  (45,\bm{1},\bm{1},\bm{1})(0) + (\bm{1},\bm{3},\bm{1},\bm{1})(0) + (\bm{1},\bm{1},\bm{3},\bm{1})(0) + (\bm{1},\bm{1},\bm{1},\bm{3})(0) + (\bm{1},\bm{1},\bm{1},\bm{1})(0)\notag\\
     &+(\bm{1},\bm{2},\bm{2},\bm{2})(2) + (\bm{1},\bm{2},\bm{2},\bm{2})(-2) + (\bm{1},\bm{1},\bm{1},\bm{1})(4) + (\bm{1},\bm{1},\bm{1},\bm{1})(-4)\notag \\
     &+(\bm{10},\bm{1},\bm{2},\bm{2})(0) + (\bm{10},\bm{2},\bm{1},\bm{1})(2) + (\bm{10},\bm{2},\bm{1},\bm{1})(-2).
\end{align}
Thus, we obtain the four-dimensional scalars that belong to $\bm{10}(0), \bm{10}(2), \bm{10}(-2)$ of $H=\mathrm{SO}(10)\times \mathrm{U}(1)$.

In this embedding, the spinors of $\mathrm{SO}(8)_\text{loc}$ transform under $R$ as
\begin{align}
    \mathrm{SO}(8)_\text{loc} \supset &~\mathrm{SU}(2)\times \mathrm{SU}(2)\times \mathrm{SU}(2)\times \mathrm{U}(1),\notag \\
    \bm{8}_\text{c} = & ~(\bm{1},\bm{2},\bm{1})(2) + (\bm{2},\bm{1},\bm{2})(0) + (\bm{1},\bm{2},\bm{1})(-2),\notag \\
    \bm{8}_\text{s} = & ~(\bm{1},\bm{1},\bm{2})(2) + (\bm{2},\bm{2},\bm{1})(0) + (\bm{1},\bm{1},\bm{2})(-2),
\end{align}
while $F=\bm{256}$ of $G=\mathrm{SO}(18)$ decomposes under $R\times H$ as
\begin{align}
    \mathrm{SO}(18) \supset &~ \mathrm{SO}(10)\times \mathrm{SU}(2)\times \mathrm{SU}(2)\times \mathrm{U}(1),\notag \\
    \bm{256} = & ~  (\bm{16},\bm{1},\bm{2},\bm{1})(2) + (\bm{16},\bm{2},\bm{1},\bm{2})(0) + (\bm{16},\bm{1},\bm{2},\bm{1})(-2) \notag \\
    &+ (\overline{\bm{16}},\bm{1},\bm{1},\bm{2})(2) + (\overline{\bm{16}},\bm{2},\bm{2},\bm{1})(0) + (\overline{\bm{16}},\bm{1},\bm{1},\bm{2})(-2).
\end{align}
If the two decompositions are compared, we obtain $\bm{16}(\pm2),\bm{16}(0)$ as $\mathrm{SO}(10)\times \mathrm{U}(1)$ representations for the left-handed spinor fields, and $\overline{\bm{16}}(\pm2),\overline{\bm{16}}(0)$ for the right-handed spinor fields. 
Taking the charge conjugation of the right-handed spinor fields, six generations of $\bm{16}$'s of $\mathrm{SO}(10)$ in total appear in the four dimensions.

From each of these three models, it is possible to obtain several generations of the fermions in the SM. 
This is because the $\bm{16}$ representation of $\mathrm{SO}(10)$ includes one generation of the SM fermions.

\section{Summary and discussion}\label{sec:summary}
In this article, we have utilized the coset space dimensional reduction (CSDR) to analyze twelve-dimensional gauge-Higgs unified models with an eight-dimensional coset space under appropriate assumptions. Then we built the twelve-dimensional models that lead to phenomenologically promising models in four-dimensional space-time. 

First, we have made a list of inputs for CSDR, including the extra space $S/R$, the twelve-dimensional gauge group $G$, and the $G$ representation $F$ of twelve-dimensional fermions. 
The eight-dimensional coset spaces $S/R$ are classified by requiring $\operatorname{rank}R=\operatorname{rank}S$, and displayed in Table \ref{tab:list_8dim_coset_modified}.
We have summarized the coset spaces shown in Table \ref{tab:list_8dim_coset_rearranged} that permit complex representations for $R$ and at most one additional $\mathrm{U}(1)$ factor in the resulting four-dimensional gauge group obtained via CSDR. 
Second, it is required that the twelve-dimensional gauge group $G$ must contain complex representations and should lead to a favorable four-dimensional gauge group $H$ that includes $G_\text{SM}$, $\mathrm{SU}(5)$, $\mathrm{SO}(10)$, and their possible extension by an extra $\mathrm{U}(1)$ factor. 
The candidates of twelve-dimensional gauge groups $G$ that satisfy all requirements have been summarized in Table \ref{tab:list_12dim_gauge_group}.
We have also limited the fermion representation $F$ to a complex representation of up to one thousand dimensions.

Then CSDR is performed for each set of inputs, and we studied the resulting particle content of the four-dimensional models. 
The twelve-dimensional models that lead to an $\mathrm{SO}(10)\times \mathrm{U}(1)$ model in four dimensions are listed in Table \ref{tab:obtained_models} as phenomenologically promising. 
Note that with our assumptions we do not find models that lead to $G_\text{SM}$, $\mathrm{SU}(5)$, or $\mathrm{E}_6$ gauge groups or their $\mathrm{U}(1)$ extensions. 
The $\mathrm{SO}(10)\times \mathrm{U}(1)$ GUT-like models include spinors in the representations $\bm{16}$ of $\mathrm{SO}(10)$ that contain one generation of the SM fermions and scalars in $\bm{10}$ which can be interpreted as Higgs fields that spontaneously break electroweak symmetry.

Among the results, we are particularly interested in the model with $S/R=\mathrm{SU}(5)\times \mathrm{SU}(4)\times \mathrm{U}(1)$, $G=\mathrm{SO}(18)$, and $F=\bm{256}$.
This model results in five generations (odd generations) of fermions that include three generations of left-handed fermions.
This is due to the triality of $\mathrm{SO}(8)$, where the $\mathrm{SO}(8)$ spinor $\bm{8}_\text{s}$ behaves like a six-dimensional vector representation of $\mathrm{SU}(4)\sim \mathrm{SO}(6)$ under $R=\mathrm{SU}(4)\times \mathrm{U}(1)$, while the other spinor $\bm{8}_\text{c}$ behaves differently from $\bm{8}_\text{s}$.
In addition, the extra $\mathrm{U}(1)$ charges for the three generations of left-handed fermions are $0$ and $\pm2$.
This charge assignment is reminiscent of the charges in the $\mathrm{U}(1)_{L_\mu-L_\tau}$ model~\cite{Foot:1990mn, He:1990pn, He:1991qd, Foot:1994vd} that has been extensively studied, for instance, in the contexts of the neutrino physics~\cite{Asai:2017ryy, Asai:2018ocx, Asai:2019ciz, Araki:2019rmw}, cosmology~\cite{Baek:2008nz, Baek:2015fea, Patra:2016shz, Biswas:2016yan, Altmannshofer:2016jzy, Arcadi:2018tly, Bauer:2018egk, Foldenauer:2018zrz, Escudero:2019gzq, Asai:2020qax, Asai:2020qlp, Araki:2021xdk, Borah:2021mri, Asai:2023wip}, and realization of the U(1)$_{L_\mu-L_\tau}$ symmetry from high-energy theory~\cite{Sato:2021fpo}.
By manipulating the two generations of right-handed fermions that appear in excess, 
we may be able to obtain left-handed fermions with $\mathrm{U}(1)_{L_\mu-L_\tau}$ charge for three generations, which could potentially explain the three-generation structure of the fermions in the SM.

After applying our currently investigated method, when a GUT model in four dimensions is obtained, this GUT symmetry must be broken down to the SM.
The scalars in the considered model are supposed to break the electroweak symmetry spontaneously and cannot be used to break the GUT symmetry.
We should consider other methods to break the GUT symmetry, for instance, via the Hosotani mechanism, also known as the Wilson flux breaking mechanism \cite{Hosotani:1983xw, Hosotani:1983vn, Hosotani:1988bm, Witten:1985xc}.

In this current study, we have taken an ansatz that only one direct product factor of the $\mathrm{U}(1)$ in $R$ is considered, just for simplicity.
Thus, if we allow for two or more $\mathrm{U}(1)$ factors which generally introduce new degrees of freedom by linear combinations of the $\mathrm{U}(1)$ generators in embedding $R$ into $G$.
It is obvious that the analysis becomes very complicated and therefore we have to find a new way to study these issues.

\section*{Acknowledgments}

This work was supported by JSPS KAKENHI Grant Numbers JP18H01210 [JS, YT, MJSY], JP21K20365 [KA], and JP23K13097 [KA], and MEXT KAKENHI Grant Number JP18H05543 [JS, YT, MJSY].

\bibliographystyle{utphys28mod}
\bibliography{csdr12v1}
\end{document}